\documentclass[prb,preprint]{revtex4}

\usepackage{latexsym}
\usepackage{amsfonts}
\usepackage{amssymb}
\usepackage{amsmath}
\usepackage{graphicx}

\begin{document}

\title{On-chip detection of ferromagnetic resonance of a single submicron permalloy strip }

\author{M. V. Costache, M. Sladkov, C. H. van der Wal and B.
J. van Wees} \affiliation{Physics of Nanodevices Group, Materials Science Centre, University of
Groningen,\\
Nijenborgh 4, 9747 AG Groningen, The Netherlands }

\date{\today}

\begin{abstract}
We measured ferromagnetic resonance of a single submicron
ferromagnetic strip, embedded in an on-chip microwave
transmission line device. The method used is based on detection
of the oscillating magnetic flux due to the magnetization
dynamics, with an inductive pick-up loop. The dependence of the
resonance frequency on applied static magnetic field agrees very
well with the Kittel formula, demonstrating that the uniform
magnetization precession mode is being driven.
\end{abstract}
\maketitle

Recent discoveries of novel phenomena in mesoscopic systems
containing nanomagnets pave the way to new spintronics
devices\cite{Kiselev,Saitoh,Tulapurkar}. \citet{brataas}
proposed a new application of ferromagnetic resonance (FMR), a
so-called spin battery. In such a device, a spin current flows
into a paramagnetic metal from its interface with a precessing
ferromagnet, which is resonantly driven with an rf magnetic
field. A number of experiments
\cite{zhang,silva,grundler_apl,ding,Celinski_01,kaka}, have been
used to measure FMR on thin ferromagnetic films or on ensembles
of small ferromagnets. This work measured rf-power transmission
or absorption with a coplanar waveguide with ferromagnetic
material on top of it, or with ferromagnetic material in a
microwave cavity. However, these measurements do not provide
enough sensitivity to measure FMR of a single submicron
ferromagnetic strip, and cannot be implemented into a lateral
multi-terminal device needed for a spin battery device.

\begin{figure}
\includegraphics[width=8cm]{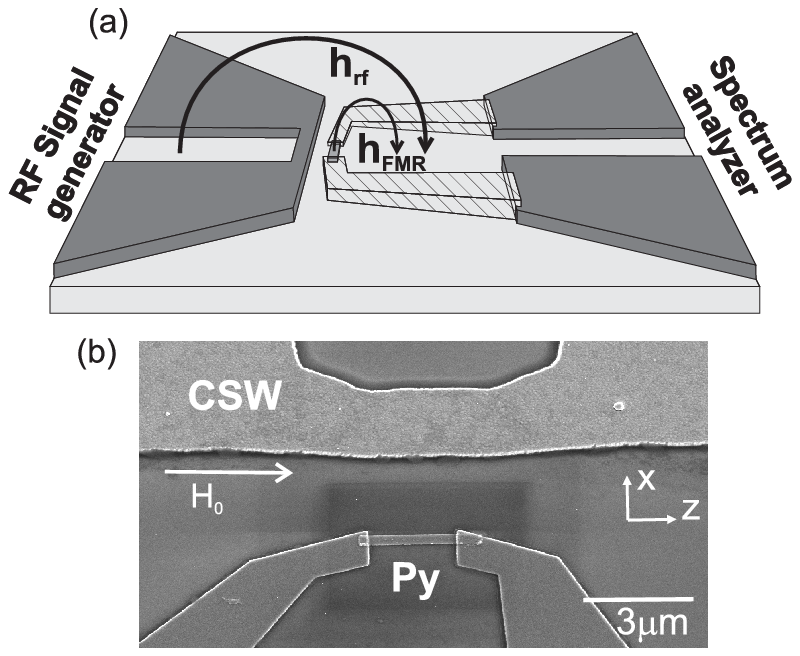}
\caption{(a) Schematic diagram of the device. On the left side
an rf current is driven through the shorted end of the coplanar
strip, which generates an rf magnetic field $h_{rf}$
(Biot-Savart law). On the right, a loop that contains a
submicron Py strip is connected to another coplanar strip. RF
power that results from oscillating magnetic flux in this loop
is measured with a spectrum analyzer. Magnetization precession
of the Py strip contributes a magnetic field $h_{FMR}$ to the
total magnetic field. (b) Scanning electron microscope picture
of central part of the device (rotated by $90^{o}$ with respect
to the upper figure).} \label{sample}
\end{figure}

We present here experiments which detect FMR of a single
submicron permalloy $\left( Py=Ni_{80}Fe_{20}\right)$ strip, and
show that driving of a $Py$ strip with an external rf field can
result in resonant excitation of pure uniform magnetization
precession (unlike earlier work on $Co$ strips \cite{julie},
where mainly domain wall resonances and possibly spin wave modes
were excited). We studied microwave transmission in a setup with
two on-chip coplanar strip wave guides (CSW), see fig.
\ref{sample}. The waveguide with the short at its end is used
for generation of a high-amplitude, localized rf magnetic field
$h_{rf}(t)$. This is used for driving the magnetization of a
submicron $Py$ strip, that is embedded in a small pick-up coil.
Oscillations of the magnetic flux in this coil induce a
microwave signal in the second CSW, of which the power is
detected with a spectrum analyzer \cite{spectrum}. Besides a
contribution from the driving field (the source flux), we find
that precession of the strip's magnetization gives a
contribution (the FMR flux) to the flux in the pick-up coil
(superposition of source flux and FMR flux), and that this can
be used as a highly sensitive probe for FMR of an individual
nanomagnet. Measurements are done by slowly sweeping a static
magnetic field $H_{0}$ applied along the $Py$ strip and
perpendicular to $h_{rf}$, while the generated power at the
applied rf frequency was measured by a spectrum analyzer
\cite{spectrum}. All the measurements are done at room
temperature.

Note, however, that symmetry arguments predict that this type of
FMR signals should be zero for our geometry. Upon precession,
the magnetization has oscillating components transverse to its
equilibrium direction (parallel to the strip's easy axis). For
these, the component in the device plane does nominally not
result in field lines that pierce the loop. For the out-of-plane
component, the flux coming out of the top surface of the strip
exactly equals the flux piercing the surrounding sample plane,
and the field falls of rapidly with increasing distance from the
strip. Consequently, also this component does not cause a flux
in the loop if the loop's boundary is assumed to coincide
exactly with the strip's central axis. Our FMR signals are due
to deviations from these exact symmetries, which are hard to
quantify for devices that we can realize at this stage. Based on
the details of our geometry (fig. \ref{sample}b), we believe
that the non-zero flux coupling is dominated by the out-of-plane
component.

The devices were fabricated by evaporating a $Py$ strip ($25~nm$
thick, $0.3\times 3$ $\mu m^{2}$ lateral size) at $2.5~ \mu m$
distance from the shorted-end of the CSW, and connecting it with
$Cu$ leads ($80~nm$ thick) to the other CSW. For $Py$ and $Cu$
e-beam lithography and lift-off was used. The CSW structures
were made of $Au$ ($300~nm$ thick) by means of optical
lithography on a lightly doped silicon wafer with a $500~nm$
thermal oxide surface layer. The two CSW are designed to have
$50~\Omega $ impedance \cite{Gupta} and connected by means of
microwave picoprobes to the rf signal generator and spectrum
analyzer. For all measurements the output power of the signal
generator was set at $20~dBm$ ($100~mW$), the power that reaches
the sample, however, is reduced by a few dB.

\begin{figure}
\includegraphics[width=8cm]{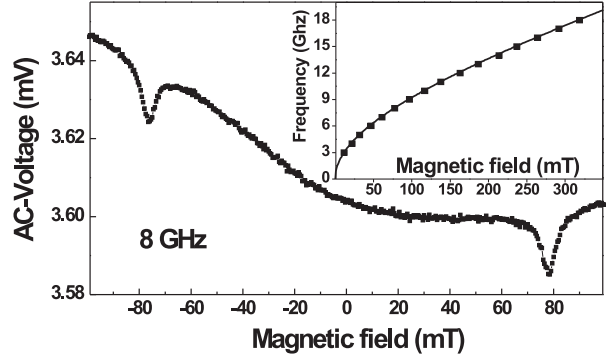}
\caption{Measured ac voltage as a function of static magnetic
field for an applied rf magnetic field at frequency $8~GHz$. The
inset shows the static magnetic field dependence of the
resonance frequency, as derived from the dips in the ac voltage
versus magnetic field. The squares represent the experimental
data, the curve is a fit to the data using the Kittel formula.}
\label{kittel}
\end{figure}

Figure \ref{kittel} shows a typical output signal as a function
of the static magnetic field ($H_{0}$), taken for an applied rf
frequency of $8~GHz$. Two well defined $20~\mu V$ dips at
$H_{0}=\pm80~mT$ are observed on top of a few $mV$ background.
By measuring the transmitted power on a chip without the
ferromagnetic strip, it was found that the background signal is
unrelated to FMR, and due to a parasitic capacitive coupling between
the two CSW (also discussed later).

The inset of Fig. 2 shows the position of the dip for different
frequencies of the rf field as a function of the static applied
field. The squares correspond to the experimental data, while
the solid line is a fit with the Kittel formula for the uniform
precession mode \cite{kittel}: $\omega _{0}^{2}=\gamma
^{2}H_{0}\left( H_{0}+M_{S}\right) $, where
$\gamma$ is the gyromagnetic ratio. From
the fit, the saturation magnetization of the $Py$ strip was
found to be about $\mu _{0}M_{S}= 1~T$ and the gyromagnetic
ratio $\gamma$= 176 $GHz/T$, consistent with
earlier reports \cite{grundler_apl}. The excellent fit
demonstrates that the magnetization vector of the $Py$ strip was
driven in the uniform precession mode.

\begin{figure}
\includegraphics[width=8cm]{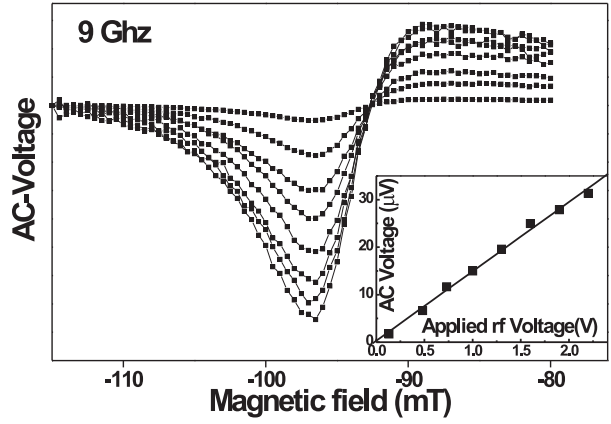}
\caption{Measured ac voltage as a function of static magnetic
field for a sequence of amplitudes of the rf driving field, at
$9~GHz$. The inset shows a linear dependence for the amplitude
of the FMR signature in the measured ac voltage on the amplitude
of the driving field (applied rf voltages scale is used, derived
from the applied power assuming $50~\Omega$ load impedance).}
\label{dip}
\end{figure}

The amplitude of the FMR signature was measured as a function of
the amplitude of the rf driving field at $9~GHz$ (fig.
\ref{dip}), showing a linear dependence. The magnetization
precession around the direction of an effective field
$\vec{H}_{eff}$ only results in a small time-dependent component
of the magnetization perpendicular to the easy axis
$\vec{M}\left( t\right) =m_{x}\left( t\right) \cdot
\hat{x}+m_{y}\left( t\right) \cdot \hat{y}+M_{S}\cdot \hat{z}$
that can be described by the linearized Landau Lifschitz Gilbert
(LLG) equation \cite{LLG}
\begin{equation}
\frac{d\vec{M}}{dt}=-\gamma \left[ \vec{M}\times
\vec{H}_{eff}\right] +\frac{\alpha }{M_{S}}\left[ \vec{M}\times
\frac{d\vec{M}}{dt}\right] \label{bigeq}
\end{equation}
where $\alpha $ is the dimensionless Gilbert damping parameter.
We solved this equation under the assumption that the strip can
be treated as a single domain thin film with a demagnetizing
field only in the out-of-plane direction and the crystal
anisotropy field is neglected, thus the effective field can be
written as $\vec{H}_{eff}=\left[ -m_{y}\left( t\right) +h\left(
t\right) \right] \cdot \hat{y}+H_{0}\cdot \hat{z}$.

The solution of this equation can be presented as the magnetic
susceptibility tensor $\chi _{ij}\left( \omega \right) $. In our
particular case only $\chi _{xy}$ and $\chi _{yy}$ components
are required and since we assume that the out-of-plane component
of magnetization dominates the magnetic flux contribution in the
loop, the $\chi_{xy}$ can be excluded from consideration while
the $\chi _{yy}$ has a form (further tensor indices are omitted
for simplicity)
\begin{equation}
\chi \left( \omega \right) =\gamma M_{S}\frac{\gamma H_{0}+i\omega
\alpha }{\omega ^{2}-\omega _{0}^{2}-i\omega \alpha \gamma \left(
2H_{0}+M_{S}\right)} \label{susc}
\end{equation}

\begin{figure}
\includegraphics[width=8cm]{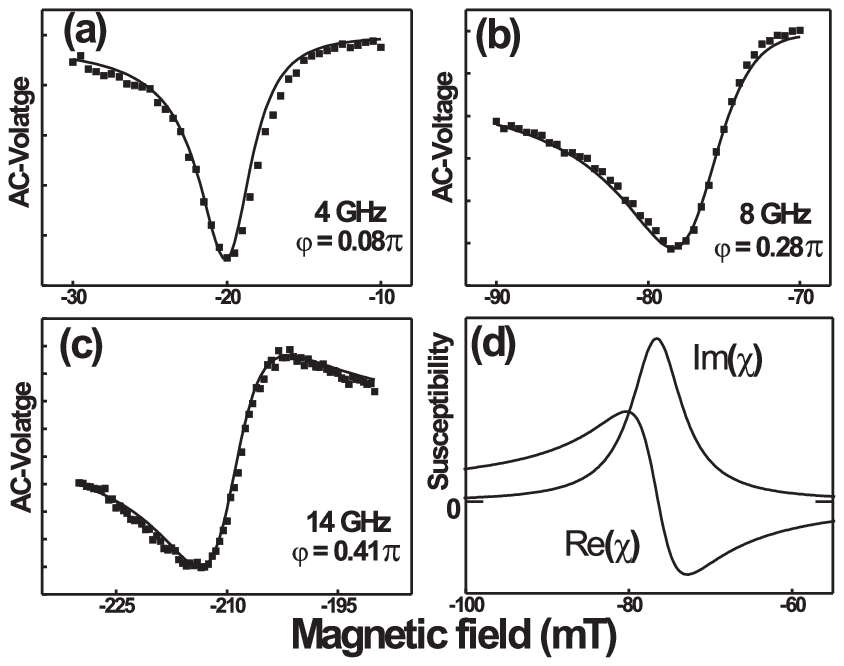}
\caption{(a),(b),(c) Measured ac voltage as a function of the
static magnetic field, around the resonance position for the
frequencies $4~GHz$, $8~GHz$ and $14~GHz$, shown by squares. The
line is the fit to the data using equation (\ref{volt}) (with
the value for $\varphi$ in the plot).  (d)The $real$ and the
$imaginary$ parts of the susceptibility $\chi$ calculated using
equation (\ref{susc}) for rf field frequency $8~GHz$,
$\gamma=176~GHz/T$, $M_{S}=1~T$ and $\alpha=0.015$.} \label{fit}
\end{figure}

In Figure \ref{fit}(d), we plot the real and imaginary parts of
magnetic susceptibility as a function of the static magnetic
field ($H_{0}$) for an rf field of $8~GHz$ frequency, with
$\gamma=176~GHz/T$, $\mu _{0}M_{S}=1~T$ and $\alpha=0.015$. The
imaginary part of $\chi \left( \omega \right) $ describes the
out-of phase magnetization precession. This results in the
absorption peak observed in conventional FMR experiments, with a
linewidth increasing linearly with frequency and being a
function of $\alpha $.

Figures \ref{fit} (a), (b), (c) show the measured signal around
the resonant frequency for three typical rf frequencies $4, 8,$
and $14~GHz$ (the squares) from a different sample. We note here
that the FMR dip shape changes from a Lorentzian to a more
complex shape as the frequency changes. This is due to a
parasitic capacitive coupling between the two CSW, meaning that
there is an extra contribution to the voltage created by the
applied rf field. This can also be understood as a phase shift
($\varphi$) between the voltage created by the source flux and
the voltage due to the FMR flux. We fit the the change in
voltage observed at resonance using the following function
\begin{equation}  \label{volt}
\Delta V( \omega ) \propto A(\omega)\cdot(Im[\chi(\omega)]\cdot
cos(\varphi) +Re[\chi(\omega )]sin(\varphi))
\end{equation}
where $A(\omega)$ depends on the amplitude of the $h_{rf}$  and
on the coupling between the time dependent magnetization and the
FMR flux generated by this. With $A(\omega)$ and $\varphi$ as
fit parameters, all the measured signals could be fit very well
Fig. \ref{fit} (a)-(c)). These fits allow us to determine the
Gilbert damping parameter, which was found to be $\alpha
=0.015$. This value is larger than the value $\alpha =0.007$
commonly accepted for a thin film of Py \cite{silva,Gerits}. Our
higher value can be due to the spatial inhomogeneities of the
driving field $h_{rf}$ and magnetic inhomogeneities at the strip
edges \cite{Celinski_01}, or due to spin pumping\cite{bauer}
from the Py strip into the Cu contacts, as previously measured
for a $Cu/Py/Cu$ film structure\cite{Ando}.

Finally, we turn to estimating the precession cone angle, an
important parameter for the spin battery proposal. An upper
limit can be estimated by assuming that all the rf power from
the signal generator leads to rf current in the short at the end
of the left CSW. The the rf field amplitude driving the strip is
then $\sim$ $3.8~mT$, which gives a precession cone angle of
$\theta _{y}\approx h_{RF}/(\alpha M_{S})=13^{o}$. A lower limit
can be obtained from the amplitude $\Delta V$ of the resonances
in the measured ac voltage, $\theta _{y}= \Delta V/(\omega \cdot
S \cdot M_{S})$, where $S$ is the effective coupling area of the
asymmetries that lead to non-zero flux coupling. If we assume an
upper limit for this asymmetry of a reasonable fraction of the
strip's surface, 3000 x 10 $nm^{2}$, a measured ac voltage
amplitude of $20~ \mu V$ at 10 Ghz gives a cone angle of $\theta
_{y}=2.5^{o}$. This indicates that the cone angle is a few
degrees in our experiment. Further examination and modelling of
the asymmetries that lead to non-zero flux coupling and the
microwave circuitry is needed for a better understanding of the
magnitude of the measured signal.

In summary, we demonstrated the resonant excitation and
detection of the ferromagnetic resonance uniform mode of a
single submicron ferromagnetic strip, embedded in an on-chip
microwave transmission line device. We obtain a precession cone
angle of a few degrees, and a Gilbert damping parameter $\alpha
=0.015$. These results are promising for further studies on new
mechanisms for controlling magnetization and electron spins in
lateral nanodevices at high frequencies, as for example the spin
battery proposal.

This work was supported by the Dutch Foundation for Fundamental
Research on Matter (FOM). We thank S. M. Watts, J. Grollier, G.
Visanescu and A. Slachter for contributions to this work, and B.
Wolfs and S. Bakker for technical support.

\newpage

\begin{center}
 \textbf{REFERENCES} \\
 \end{center}

\newpage

\newpage
\clearpage

\newpage
\clearpage

\newpage
\clearpage

\begin{thebibliography}{18}
\bibitem[{spe()}]{spectrum}
\bibinfo{note}{The spectrum analyzer measures power, which we convert in a
  voltage scale assuming $50~\Omega $ load impedance. To improve the
  signal-to-noise ratio we employ a frequency locked technique with a 10 Hz
  spectral bandwidth}.
\end{thebibliography}
\end{document}